\documentclass[aip,jap,reprint,superscriptaddress,float,amsmath]{revtex4-2}
\usepackage{xcolor}
\usepackage{graphicx}
\usepackage{amsmath,epsfig}
\definecolor{g-blue}{rgb}{0.83,0.95,1}
\usepackage{bm}
\usepackage{amssymb}
\usepackage{amsfonts}
\usepackage{amsmath}
\usepackage{graphicx}
\usepackage{ulem}
\usepackage{url}
\usepackage{textcomp} 
\usepackage{siunitx}

\usepackage[pdftex,colorlinks=true,allcolors=blue,breaklinks=true]{hyperref}  

\makeatletter
\def\set@curr@file#1{%
	\begingroup
	\escapechar\m@ne
	\xdef\@curr@file{\expandafter\string\csname #1\endcsname}%
	\endgroup
}
\def\quote@name#1{"\quote@@name#1\@gobble""}
\def\quote@@name#1"{#1\quote@@name}
\def\unquote@name#1{\quote@@name#1\@gobble"}
\makeatother

\newcommand{\etal}{\textit{et al.~}}

\begin{document}
	
	\title{Confinement of Bose-Einstein magnon condensates in adjustable complex magnetization landscapes}
	 
	\author{Matthias~R.~Schweizer}
	\email{mschweiz@rhrk.uni-kl.de}
	\affiliation{Fachbereich Physik and Landesforschungszentrum OPTIMAS, Technische Universit\"{a}t Kaiserslautern, 67663 Kaiserslautern, Germany}
	
	\author{Alexander~J.~E.~Kreil}
	\affiliation{Fachbereich Physik and Landesforschungszentrum OPTIMAS, Technische Universit\"{a}t Kaiserslautern, 67663 Kaiserslautern, Germany}
	
	\author{Georg~von~Freymann}
	\affiliation{Fachbereich Physik and Landesforschungszentrum OPTIMAS, Technische Universit\"{a}t Kaiserslautern, 67663 Kaiserslautern, Germany}
	\affiliation{Fraunhofer-Institute for Physical Measurement Techniques IPM, Fraunhofer-Platz 1, 67663 Kaiserslautern, Germany}

	\author{Burkard~Hillebrands}
	\affiliation{Fachbereich Physik and Landesforschungszentrum OPTIMAS, Technische Universit\"{a}t Kaiserslautern, 67663 Kaiserslautern, Germany}

	\author{Alexander~A.~Serga}
	\email{serga@physik.uni-kl.de}
	\affiliation{Fachbereich Physik and Landesforschungszentrum OPTIMAS, Technische Universit\"{a}t Kaiserslautern, 67663 Kaiserslautern, Germany}
	
	\date{\today}
	
\begin{abstract}

Coherent wave states such as Bose-Einstein condensates (BECs), which spontaneously form in an overpopulated magnon gas even at room temperature, have considerable potential for wave-based computing and information processing at microwave frequencies. The ability to control the transport properties of magnon BECs plays an essential role for their practical use.
Here, we demonstrate spatio-temporal control of the BEC density distribution through the excitation of magnon supercurrents in an inhomogeneously magnetized yttrium iron garnet film. The BEC is created by microwave parametric pumping and probed by Brillouin light scattering spectroscopy. The desired magnetization profile is prepared by heating the film with optical patterns projected onto its surface using a phase-based wavefront modulation technique. 
Specifically, we observe a pronounced spatially localized magnon accumulation caused by magnon supercurrents flowing toward each other originating in two heated regions. This accumulation effect increases the BEC lifetime due to the constant influx of condensed magnons into the confinement region.
The shown approach to manipulate coherent waves provides an opportunity to extend the lifetime of freely evolving magnon BECs, create dynamic magnon textures, and study the interaction of magnon condensates formed in different regions of the sample.
\end{abstract}

\maketitle

%
%
%
\section{Introduction}
Bose-Einstein condensation is a macroscopic quantum phenomenon, which occurs spontaneously at thermodynamic equilibrium in a bosonic many-particle system at low temperatures or high particle densities. The phenomenon manifests itself through the emergence of a coherent state of matter, consisting of a macroscopic number of particles with the same energy and momentum at the lowest energy level of the system.
Its existence was predicted for an ideal bosonic quantum gas by Satyendra Nath Bose and Albert Einstein in 1924-1925. \cite{Einstein2005} In 1968, Herbert Fr\"{o}hlich proposed the concept of condensation of electric modes driven out of thermal equilibrium by external excitation. \cite{Frohlich1968} While the Bose-Einstein condensate (BEC) closest to the ideal gas model was first observed in low-density clouds of ultra-cold atoms, \cite{Anderson1995, Davis1995} well-known phenomena such as superfluidity in liquid $^4 \mathrm{He}$ and superconductivity of Cooper pairs can also be related to Bose-Einstein condensation. Later, equilibrium and non-equilibrium  BEC-like \cite{Rodrigues2018} states have been found in various quasi-particle systems such as magnons in liquid $^3 \mathrm{He}$ \cite{BorovikRomanov1984, Bunkov2007} and solid-state ferromagnets \cite{Demokritov2006, Serga2014, Safranski2017, Schneider2020, Divinskiy2021}, triplons in dimerized quantum antiferromagnets, \cite{Giamarchi2008} exciton-polaritons, \cite{Kasprzak2006, Lerario2017} phonons, \cite{Rodrigues2006}  and photons \cite{Klaers2010, Damm2016}. In many of the quasiparticle systems, supercurrent and superfluid effects, \cite{Borovik-Romanov1988, Volovik2008, Amo2009, Bozhko2016} and related phenomena such as Josephson oscillations, \cite{Lagoudakis2010, Abbarchi2013, Kreil2021, Autti2020} Bogoliubov waves, \cite{Bozhko2019} and quantized vorticity \cite{Nowik-Boltyk2012} were reported.
Their use in magnonics \cite{Pirro2021, Barman2021, Chumak2022} and spintronics \cite{Hoffmann2015} opens the way to a new generation of functional and logical devices that exploit the spontaneous coherence of Bose-Einstein condensates and their unusual transport properties. \cite{Dzyapko2008, Rezende2009, Nakata2015, Safranski2017, Tserkovnyak2017, Bunkov2020, Noack2021a, Mohseni2022} The essential task is to control the characteristics of magnon condensates.

This work focuses on the employment of supercurrents as a means of transport inside of a magnon BEC. The BEC is observed in an in-plane magnetized monocrystalline films of yttrium iron garnet (YIG), a dielectric ferrimagnet with very low magnetic damping. \cite{Cherepanov1993, Chumak2017} In this system, condensation can be quite easily accomplished by overpopulating the magnon gas by parametric electromagnetic pumping. Furthermore, the high, experimentally achievable magnon densities, reaching $10^{19} - 10^{20} \mathrm{cm}^{-3}$, allow the realization of Bose-Einstein condensation conditions even at room temperature. \cite{Demokritov2006, Serga2014} 
As the density of the gas increases, its chemical potential rises to the bottom of the magnon spectrum, and two energy-degenerated BEC states with opposite wavevectors $\pm q_\mathrm{BEC}$ and zero group velocities are formed there. \cite{Mohseni2022}
  

The straightforward approach for spatial manipulation of a BEC is to control the surrounding energy landscape, for instance by applying an additional, spatially confined bias magnetic field. It has been shown that this kind of artificial topology can lead to redistribution of the magnon density inside the condensate \cite{Borisenko2020, Borisenko2020_2, Kreil2021}. However, since the local magnetic fields are created by electric currents flowing through strip conductors on the surface of a YIG film, this method comes with demanding restrictions in terms of landscape design. 

On the contrary, changing the saturation magnetization opens up an alternative path. It has been shown that efficient control of the magnon energy landscape and realization of supercurrent BEC transport can be achieved using optical heating. \cite{Bozhko2016, Kreil2018, Bozhko2019} An increased temperature results in a decreased saturation magnetization, which in turn leads to a lowered frequency of the condensate. Although the action of this method on the BEC frequency is weaker and slower compared to the manipulation of the magnetization field, it provides an opportunity to investigate the effect of diverse form factors of reconfigurable energy landscapes \cite{Vogel2015, Vogel2018} on the dynamics of the magnon BEC.

Here, using the optical heating method, we achieved an organized spatial redistribution of the BEC density caused by the flux of condensed magnons from heated to cold regions of the YIG film. In particular, we show that by choosing an appropriate distance between the two higher temperature regions, it is possible to form a pronounced BEC occupancy peak between them, and to extend the condensate lifetime in such a heat trap.

\section{Experimental setup}
The experimental setup shown in Fig.\,\ref{F:Setup} can be divided into three modules: The excitation module for the generation of the condensate via parallel parametric pumping, the optical heating module for forming and projecting optical heating patterns onto the YIG film sample, and the optical detection module using Brillouin light scattering spectroscopy (BLS). The aim of the following sections is to describe the function of each of these modules, and subsequently, their interplay. For the control of each individual module, the automation framework \textit{thaTEC:OS} (THATec Innovation GmbH) was used. The subsequent data evaluation was performed using \textit{Python} along with additional libraries such as \textit{PyThat} and \textit{xarray}. \cite{THATec, PyThat, hoyer2017xarray, 2020SciPy-NMeth}

\subsection{Microwave-setup for magnon excitation}
The excitation module is designed to create magnons using a parallel parametric pumping process.
In this process, the microwave pumping magnetic field $\boldsymbol{h}_\mathrm{p}$ is directed parallel to the external, constant magnetization field $\boldsymbol{H}_\mathrm{ext}$ and, therefore, along the equilibrium direction of the magnetization vector $\boldsymbol{M}$. Due to the ellipticity of the precession motion of $\boldsymbol{M}$, the length of its longitudinal component, oriented along the field $\boldsymbol{H}$, is not conserved but oscillates with twice the precession frequency. When the pumping frequency coincides with this oscillation frequency,  energy transfer to the spin system occurs. \cite{Bracher2014, Schlomann1960} In this case, a microwave photon of the pumping field with a near-zero wavenumber $q_\mathrm{p} \approx 0$ decays into two magnons of half the pumping frequency $\omega_\mathrm{p}/2$ with oppositely directed wave vectors $\pm q_\mathrm{pm}$. 

When the pumping power $P_\mathrm{p}$ exceeds the threshold of parametric instability,\cite{Mihalceanu2018} that is, after the influx of parametrically generated magnons exceeds their losses, the magnon number begins to grow exponentially until it saturates, since the nonlinear phase mismatch between the electromagnetic pumping and the longitudinal magnetization oscillations limits further amplification. \cite{Lvov1994book} 
Not limited by magnetic losses, this process provides an efficient magnon injection mechanism required for Bose-Einstein condensation. The injected magnons move to the bottom of the spectrum due to cascading four-magnon scattering and kinetic instability processes \cite{Bozhko2015, Clausen2015, Kreil2018}. At a pump power of about 24\,dB above the threshold of the parametric instability, the threshold of BEC formation is reached, and a strong magnon accumulation in a narrow phase volume near the lowest energy state $(\omega_\mathrm{BEC}, \pm q_\mathrm{BEC})$ occurs. \cite{Noack2021a}
It should be noted that in our experiment, as in several previous works \cite{Serga2014, Kreil2018, Kreil2019, Noack2021a}, the formation of a coherent magnon condensate evolved during the free evolution of the congested magnon gas only after the pumping action was turned off.

\begin{figure}[t]
	\includegraphics[width=1.0\linewidth]{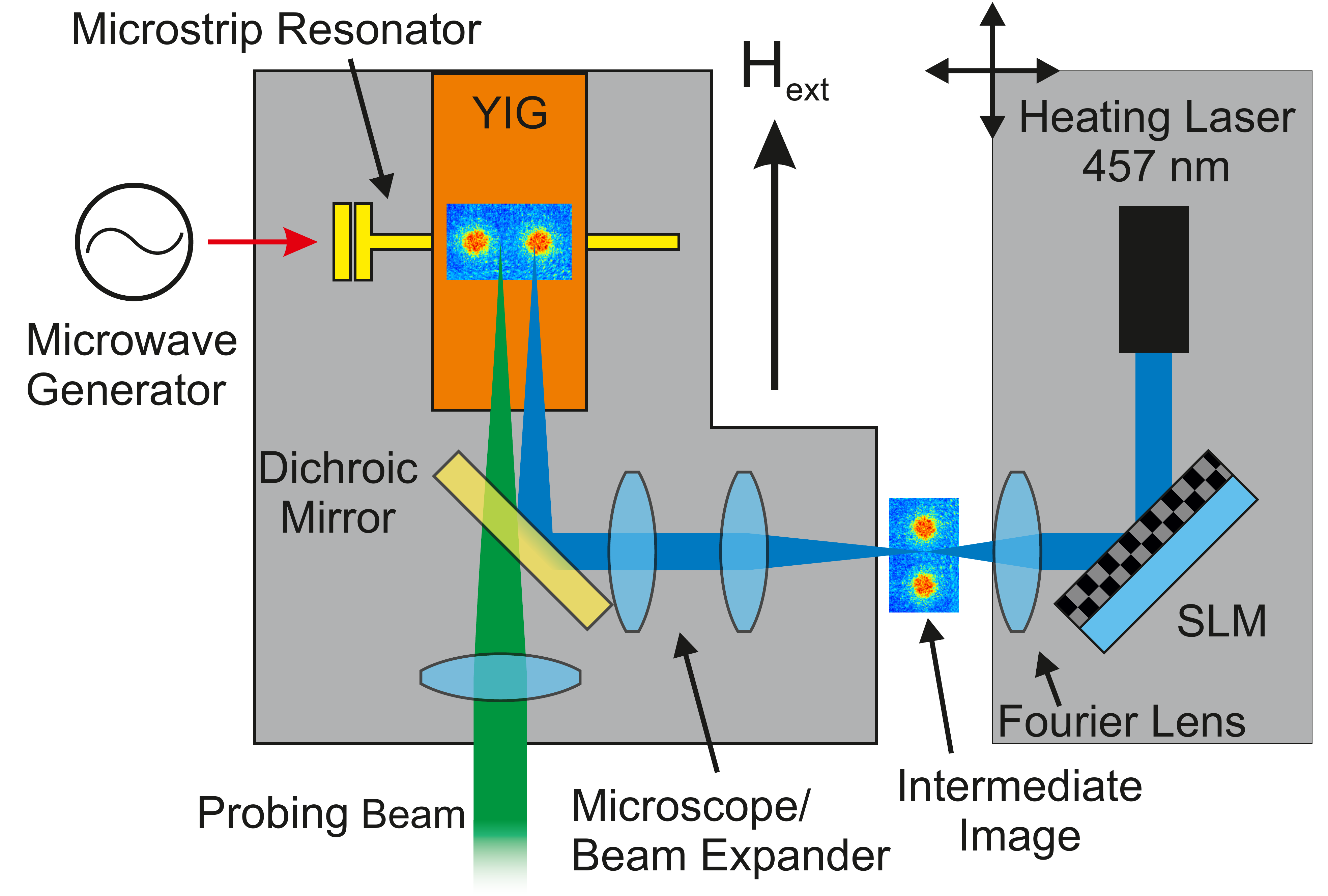}
	\caption{Schematic depiction of the experimental setup. The optical heating module is shown on the right, the YIG sample and the BLS probing beam in the middle and the microwave components on the left. The optical heating module is fixed on a motorized stage, which offers the possibility to move the intermediate image and therefore the projection of the heating pattern on the sample surface.}
	\label{F:Setup}
\end{figure}

In our experiment (see Fig.\,\ref{F:Setup}), the YIG film sample is magnetized by a bias magnetic field $\boldsymbol{H}_\mathrm{ext}$ of \SI{180}{\milli\tesla}. The pumping field $\boldsymbol{h}_\mathrm{p}$ is induced by a microwave electric current in a \SI{100}{\micro\meter}-wide  microstrip resonator tuned to the pumping frequency \hbox{$\omega_\mathrm{p}=2 \pi \cdot \SI{14}{GHz}$} and placed in direct contact with the YIG film of \SI{5}{\micro\meter}-thickness.
The resonator is fed by microwave pulses of \SI{1}{\micro\second} duration and \SI{40}{\watt} power with a repetition frequency of \SI{1}{\kilo\hertz} from a microwave generator and power amplifier.
The chosen pumping parameters allowed the formation of BECs after the end of the pump pulses but did not lead to appreciable heating of the YIG film by relaxing magnons and microwave currents.  

\subsection{Heating module for spatial supercurrent manipulation}
The heating module is the core piece for the control of the magnon BEC. It creates reconfigurable energy landscapes in which the dynamics of a BEC can be observed. By projecting complex optical intensity landscapes onto the surface of the sample, the local temperature can be increased. In consequence, the saturation magnetization $M_\mathrm{s}$ is decreased, which results in a local drop of the spin wave frequency spectrum \cite{Bozhko2019, Kreil2018, Vogel2015, Vogel2018}.

These optical intensity landscapes were created by means of phase-based wavefront modulation of light. This was achieved by using a spatial-light-modulator (SLM) \textit{Santec SLM-100}, a 2D liquid crystal display, which can imprint a phase map onto a coherent beam of light by changing the refractive index of each pixel individually. The phase-maps used for the wavefront modulation are calculated with a Gerchberg-Saxton algorithm \cite{Gerchberg1972, Vogel2015, Vogel2018, Alsaka2018, Wang2016}. An intermediate image of the desired intensity pattern is then created by placing a lens at its focal distance behind the SLM (see Fig.\,\ref{F:Setup}). This lens creates an image of the Fourier transform of the previously calculated phase map, which in turn corresponds to the desired intensity pattern. An iris aperture was added at this position, blocking higher orders of the diffraction pattern created by the SLM. The intermediate image is then demagnified by a microscope lens-system, decreasing the size of the projected image on the sample. 

Since the optical absorption of YIG increases drastically when approaching the UV spectrum\cite{Doormann1984}, a laser source \textit{Cobolt Twist}, which operates at \SI{457}{\nano\meter}, was chosen for the illumination of the SLM. This way, the optical intensity is almost completely absorbed even by thin YIG-layers. Furthermore, the small line-width of the light source and hence long coherence length grant high contrast and resolution of the resulting image.
The thermal landscape was created by continuous laser heating with \SI{20}{\milli\W} of power at the sample. In order to erase the previous structure before the start of a new measurement, the laser heating was disabled for one minute, whenever the heating pattern had been altered.

\subsection{BLS-Module for BEC detection}
The detection of the magnon density at the bottom of the spin-wave spectrum is achieved by means of frequency-, wavevector-, time-, and space-resolved Brillouin light scattering (BLS) spectroscopy \cite{sandercock1975, Sandweg2010, Bozhko2020, Buttner2000}. In this method, the intensity of light inelastically scattered by magnons is proportional to the density of magnons involved in the process, and the frequency shift of light is equal to the frequency of these magnons.
The direction of the external magnetic field is parallel to the projection of the probing beam on the sample surface, which allows for the detection of dipole-exchange magnons in the \textit{backward volume} geometry. \cite{Rezende2020} The magnon wavevectors can be selected with resolution $\pm$ \SI{0.2}{\radian\per\micro\meter} by changing the angle of incidence of the probing light. \cite{Sandweg2010, Bozhko2020, Bozhko2017, Bozhko2019}. The probing laser source is a \textit{Coherent Verdi} laser with a wavelength of \SI{532}{\nano\meter} under an angle of incidence of \SI{11}{\degree}, which corresponds to a detected magnon wavenumber of \SI{4.5}{\radian\per\micro\meter}. The area of detection is determined by the diameter of the laser spot of \SI{50}{\micro\meter}. In order to prevent the heating of the sample due to the optical power induced by the probing laser beam, this beam is switched off using an acousto-optic modulator as soon as the magnon condensate relaxes and is switched on just before the next pumping pulse is delivered. The low repetition rate, along with a short measurement window of a few microseconds, ensures that the average optical power is negligible.

\begin{figure}[tb]
	\includegraphics[width=1.0\linewidth]{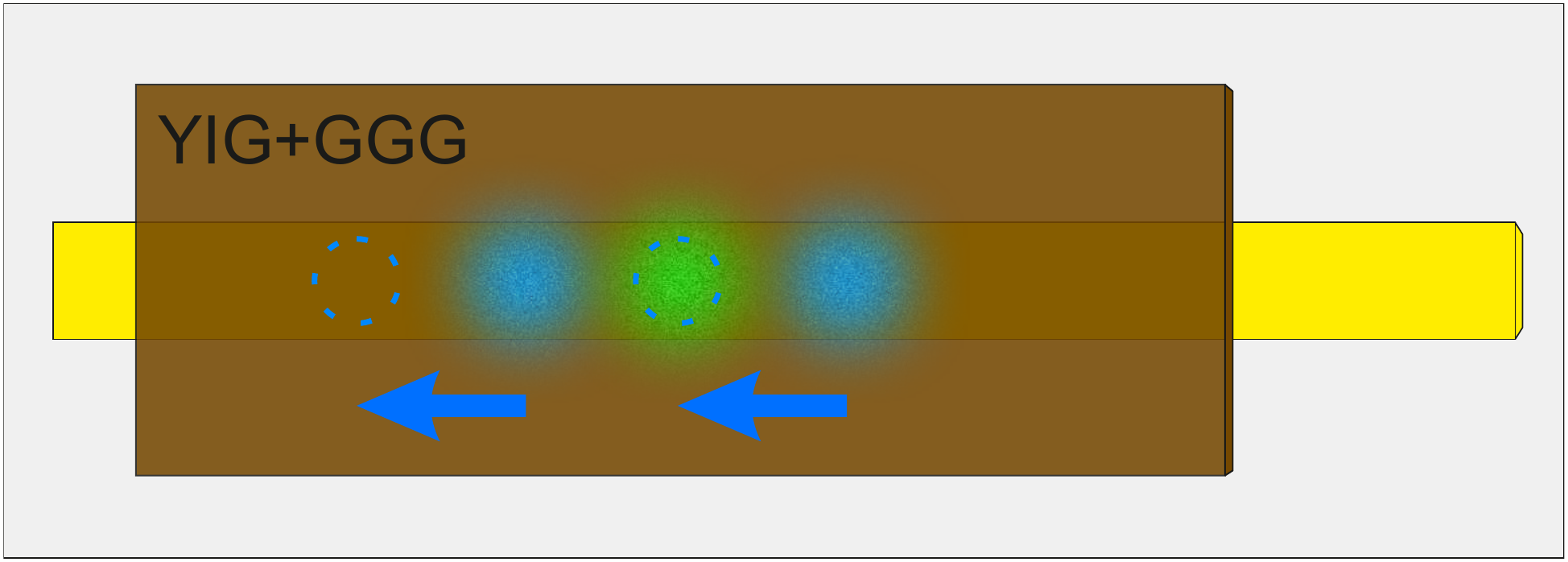}
	\caption{Schematic of the configuration of motion. The filled green and blue circles represent the initial position of the probing spot and the two heating spots, respectively. The dotted blue circles depict the displacement of the heating laser pattern. The probe spot remains in a static position while the heating pattern moves along the antenna, as shown by blue arrows.}
	\label{F:movement}	
\end{figure}

\subsection{Combination of BLS and SLM}
In order to simultaneously illuminate the sample with the probing beam of the detection module and project an optical intensity landscape with the heating module, a dichroic long-pass mirror with a reflectivity cut-off wavelength of \SI{490}{\nano\meter} is placed directly above the sample (Fig.\,\ref{F:Setup}). While the probing beam can pass through this mirror unperturbed, the heating beam is reflected, creating an additional port for optical coupling. The heating module is mounted on a separate, motorized table, whose position can be controlled independently. The intermediate image created by the Fourier lens serves as an interface between the BLS module and the heating module.

\begin{figure*}[t]
	\includegraphics[width=1.0\linewidth]{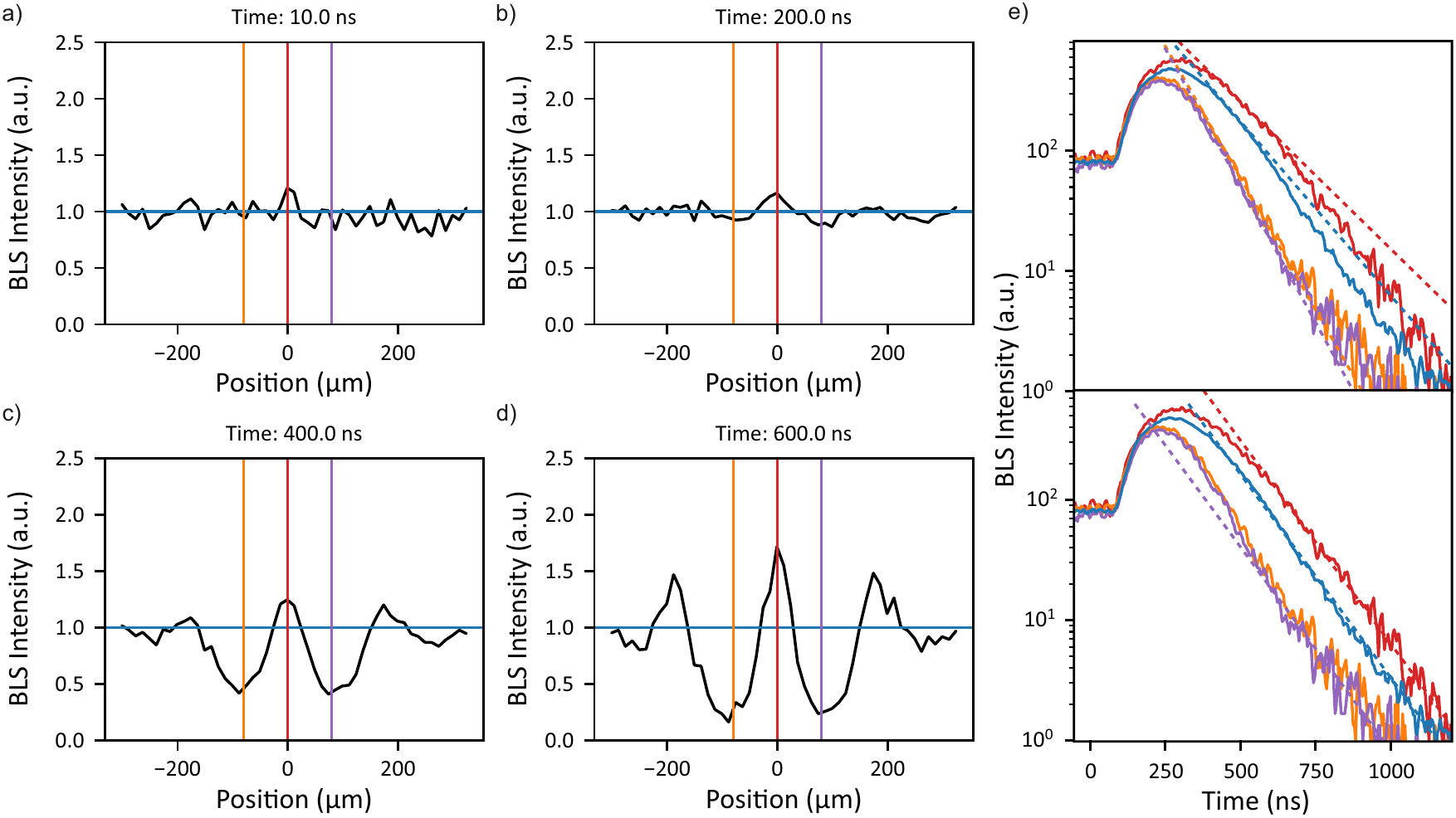}
	\caption{The four panels on the left side show cross sections of the integrated BLS intensity in the vicinity of the two-spots heating pattern with an inter-spot-distance of \SI{160}{\micro\meter}. The cross sections are depicted for a) \SI{10}{\nano\second} b) \SI{200}{\nano\second} c) \SI{400}{\nano\second} d) \SI{600}{\nano\second} after switching off the pumping pulse. The intensity has been normalized to the corresponding reference signal without heating, shown by the blue line. The time traces as shown in e) are taken at different positions, marked by the red (between the spots), orange (left spot), purple (right spot), and blue (reference) markings. The dotted lines show the regression of an exponential decay: In the upper panel, the time interval between \SI{300}{\nano\second} and \SI{550}{\nano\second} has been used, in the lower panel the interval between \SI{550}{\nano\second} and \SI{1200}{\nano\second}. }
	\label{F:Data1}	
\end{figure*}

Since the probing beam and the heating beam are mechanically decoupled, the intermediate image of the optic intensity pattern can be moved freely. This offers the possibility to move the intensity pattern over the sample while the location of investigation may remain stationary (see Fig.\,\ref{F:movement}). In this case, the magnification factor of the microscope connecting both modules has to be considered as a scaling factor between the movement of the stage and the movement of the resulting intensity pattern. Note that, since the probing point is fixed and only the structure of the sample changes, the reference case without heating is represented by only one spatial position.

Although the described wavefront modulation technique can be used to create very complex intensity patterns \cite{Wang2016, Vogel2015, Vogel2018}, in this study, we focused on a simple two-point structure with a varying distance. On the one hand, the chosen configuration represents the most fundamental case of a periodic structure. On the other hand, concentrating the available optical power in a small spatial area maximizes the heating effect on the BEC behavior. Since the magnon condensate is located above the microstrip antenna, \cite{Kreil2021, Serga2012} the chosen optical pattern also fits the system's symmetry. Therefore, the probing beam and the heating pattern were accurately positioned over the microstrip resonator before each measurement.

\section{Experimental Results}

\subsection{Spatial distribution of the magnon condensate}
Figure\,\ref{F:Data1} shows the measured BLS-intensity as a function of time and space. The data is integrated over the frequency interval between \SI{4.0}{\giga\hertz} and \SI{4.6}{\giga\hertz}. Panels (a), (b), (c), and (d) show the resulting magnon density as a function of the relative position of the heating pattern for different moments in time. While no influence of the temperature landscape is visible right after the end of the pumping pulse at $t=0$ (see Fig.\,\ref{F:Data1}a), the formation of two pronounced dips can be observed, starting at around \hbox{\SI{200}{\nano\second} (b)}. Simultaneously a pronounced peak emerges between both dips along with two weaker peaks at the outer edges of the heating spots at \hbox{$y=\pm$\SI{80}{\nano\meter}}. Although the absolute magnitude of the integrated BLS intensity decreases exponentially with time due to inherent damping of the freely evolving magnon system, the contrast between the dips and peaks only gets more pronounced. And while the magnon density far away from the heated region is comparable to the reference case depicted by the horizontal blue line, the BLS intensity between both intensity spots is up to two times as high. 
It can be assumed that---although not directly exposed to optical heating---the region between the two heating spots is still significantly hotter than in the case without any optical heating. 
This suggests that the difference in magnon density between heated and unheated regions is caused not just by the difference in their temperatures but by the presence of a specific temperature gradient.  
Although increasing the temperature of the YIG film may reduce the efficiency of the BLS scattering process, \cite{Olsson2018} it does not explain the increased BLS intensity between the heating spots.

On a side-note, in Figs.\,\ref{F:Data1}c-\ref{F:Data1}d, there are also two more intensity dips at \hbox{$\pm$ \SI{240}{\micro\meter}}. These are due to higher order diffraction orders in the intensity landscape, which appear with about \SI{10}{\percent} of the intensity of the central order. 
\subsection{Temporal dynamics of the magnon condensate}
The temporal behavior of the magnon population in the heated and cold regions in comparison is depicted in Fig.\,\ref{F:Data1}e. Both regions show the same BLS intensity during pumping and even at the beginning of the condensation process until \SI{200}{\nano\second}. However, at the peak of the condensation process, \cite{Serga2014} the behavior of the magnon populations in these regions begins to differ. At this point, the magnon density in the heated regions begins to decrease significantly faster than in the reference case. In contrast, the magnon population in the region between the two heating spots decreases significantly more slowly than in the heated regions, even more slowly than in the reference situation (see the upper panel). This difference persists until about \SI{550}{\nano\second}, when the decay rates begin to equalize again (see lower panel).

As a result, it stands to reason that the characteristic of the magnon distribution is caused by magnon transport from hot to the cold regions. While a continuous outflow of magnons from the heated region accelerates the local effective decay, the opposite effect occurs in neighboring colder regions, where an influx of magnons works against the internal attenuation.

The observed phenomenon can be associated with the emergence of magnon supercurrents as found by Bozhko \etal \cite{Bozhko2016, Bozhko2019}.  In this work, a large decrease in the magnon density in the heated region, caused by the outflow of the magnon BEC, was also recorded.

\subsection{BEC behavior as a function of the inter-spot-distance}
\begin{figure}[t]
	\includegraphics[width=1.0\linewidth]{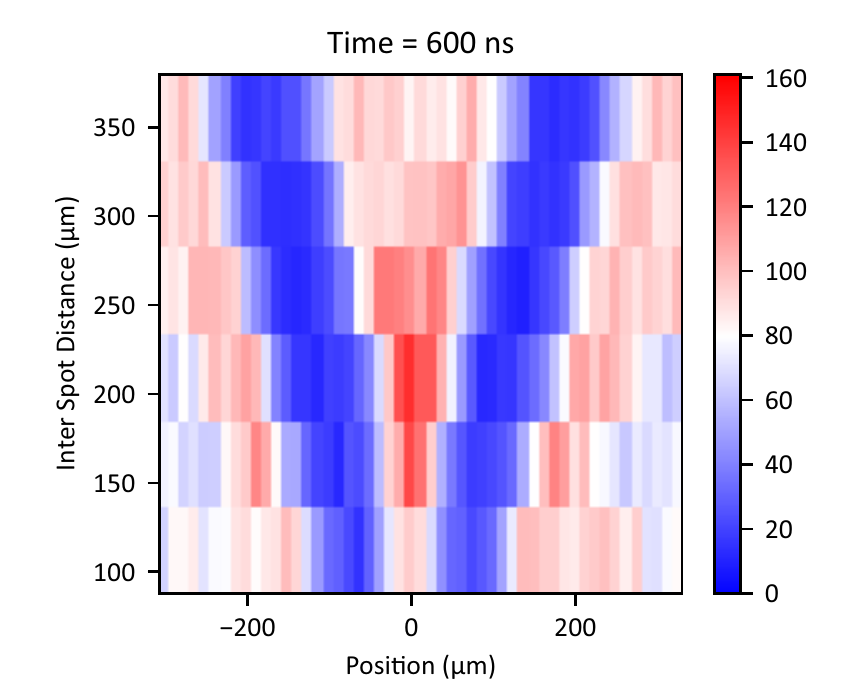}
	\caption{Color-map of cross-sections at different inter-spot-distances. The color code has been chosen in a way that white refers to values of the reference signal, while red depicts more and blue less intensity. The distance between the two heated regions, is varied between \SI{112}{\micro\meter} and \SI{355}{\micro\meter}. For all observed distances, two local minima and a local maximum of the magnon density could be observed between both heated regions. While the accumulation is most pronounced for \SI{160}{\micro\meter}, it is weakened for large or very small distances.}
	\label{F:Data2}	
\end{figure}

Figure\,\ref{F:Data2} ultimately shows the central benefit of the measurement technique presented above. The distance between the two heating spots can be changed without changing any other parameter, including the position of the probing point and the location of the sample in the magnetization field. As for the data shown in Fig.\,\ref{F:Data1}, the position of the temperature landscape was moved, whilst the position of the probing beam remained static. 

In Fig.\,\ref{F:Data2}, the deviation from the mean BLS intensity as a function of position and inter-spot distance is shown in a false color scheme. For all of the investigated inter-spot-distances a similar situation is observed. The magnon density drops strongly in the heated region, whereas it is increased above the reference level in the adjacent colder regions. However, although the magnitude of the intensity dips is comparable for all situations studied, the magnon density between the heating spots increases to a lesser degree for larger distances. 
This behavior can be attributed to a dilution effect: As long as the system characteristics responsible for the spatial transport of magnons remain constant, the total number of transported magnons is also constant. As the inter-spot-distance increases, these magnons are spread between the two spots, which leads to a decrease in the observed BEC density. However, for the shortest investigated distance, the two high-temperature areas begin to overlap due to thermal diffusion. This weakens the temperature gradient toward the middle point and, consequently, the transport of magnons. So, the contrast in the magnon density is also weakened in this configuration. In this case, thermal diffusion during continuous heating can be recognized as a limiting factor for optical resolution and feature size.

\section{Conclusion}
In conclusion, we were able to utilize magnon supercurrents to achieve trapping of a magnon Bose-Einstein-condensate at a predefined position. The continuous magnon flux towards the trapped region results in increased lifetime of the condensate in that area. Consequently, the density of the condensate can be controlled by changing the spacing between the adjacent regions with decreased saturation magnetization. Under optimal conditions, an increase in density of up to 100\,\% could be achieved. It has been shown, that the demonstrated method offers the potential to create complex magnetization landscapes, which would be challenging to create by other means.

\begin{acknowledgments} 
Funded by the Deutsche Forschungsgemeinschaft (DFG, German Research Foundation)---TRR\,173/2---268565370 (Project B04).
\end{acknowledgments}



\bibliography{two-spots}
\end{document}